\documentclass[]{aa}

\usepackage{graphicx}
\usepackage{rotating}
\usepackage{txfonts}
\usepackage{color}
\usepackage{amsmath}

\usepackage{float}
\usepackage{caption}
\usepackage{subcaption}
\usepackage{hyperref}
\usepackage{ulem}
\usepackage{comment}
\usepackage{arydshln}

\hypersetup{colorlinks=true,linkcolor=blue,citecolor=blue,urlcolor=blue}

\newcommand\ha{H$\alpha$}

\newcommand\hb{H$\beta$}

\newcommand\pab{Pa$\beta$}
\newcommand\Pab{Paschen $\beta$}
\newcommand\brg{Br$\gamma$}
\newcommand\Brg{Brackett $\gamma$}

\newcommand\Msunyr{$M_{\odot}\,{\rm yr^{-1}}$}
\newcommand\kms{${\rm km\,s^{-1} }$}
\newcommand\Msun{$M_{\odot}$}
\newcommand\Rsun{$R_{\odot}$}

\begin{document}

    \title{Spectroscopic and interferometric signatures of magnetospheric accretion in young stars} 
    \titlerunning{Spectroscopic and interferometric signatures of magnetospheric accretion}

   \author{{\sc B.~Tessore}\inst{1}, {\sc A.~Soulain}\inst{1}, {\sc G.~Pantolmos}\inst{1}, {\sc J.~Bouvier}\inst{1}, {\sc C.~Pinte}\inst{1,2}, and \sc K.~Perraut\inst{1}}
   \institute{Université Grenoble Alpes, CNRS, IPAG, 38000 Grenoble, France
   \and
   School of Physics and Astronomy, Monash University, VIC 3800, Australia
   }

   \authorrunning{B.~Tessore \& A.~Soulain et al.}

   \date{xx/xx/xx; yy/yy/yy}


  \abstract
   {}
   {We aim to assess the complementarity between spectroscopic and interferometric observations in the characterisation of the inner star-disc interaction region of young stars. }
   {We use the code MCFOST to solve the non-LTE problem of line formation in non-axisymmetric accreting magnetospheres.
   We compute the \brg{} line profile originating from accretion columns for models with different magnetic obliquities. We  also derive monochromatic synthetic images of the \brg{} line emitting region across the line profile. This spectral line is a prime diagnostics of magnetospheric accretion in young stars and is accessible with the long baseline near-infrared interferometer GRAVITY installed at the ESO Very Large Telescope Interferometer. } 
   {We derive \brg{} line profiles as a function of rotational phase and compute interferometric observables, visibilities and phases, from synthetic images. The line profile shape is modulated along the rotational cycle, exhibiting inverse P Cygni profiles at the time the accretion shock faces the observer. The size of the line's emission region decreases as the magnetic obliquity increases, which is reflected in a lower line flux. We apply interferometric models to the synthetic visibilities in order to derive the size of the line-emitting region. We find the derived interferometric size to be more compact than the actual size of the magnetosphere, ranging from 50 to 90\% of the truncation radius. Additionally, we show that the rotation of the non-axisymmetric magnetosphere is recovered from the rotational modulation of the \brg-to-continuum  photo-centre shifts, as measured by the differential phase of interferometric visibilities.
   }
   {
   Based on the radiative transfer modelling of non-axisymmetric accreting magnetospheres, we show that simultaneous spectroscopic and interferometric measurements provide a unique diagnostics to determine the origin of the \brg{} line emitted by young stellar objects and are ideal tools to probe  the structure and  dynamics of the star-disc interaction region.
   }


   \maketitle
%

\section{Introduction}
The early evolution of low mass stars ($M_{\ast}$ < 2~\Msun{}) during the classical T~Tauri (CTT) phase depends on the interaction between the star and its accretion disc, on a distance of a few stellar radii.
At the truncation radius, matter from the disc surface is channelled onto the stellar surface following the magnetic field lines and forming an accretion funnel or column \citep{Ghosh77,Zanni09,Romanova16_book,Pantolmos20}. The star-disc interaction is responsible for accretion and ejection phenomena that have a strong impact on spectral lines formed in the close vicinity of the star's surface.

\cite{Ghosh77} developed an analytical model of magnetospheric accretion around a rotating neutron star with a dipolar magnetic field. \cite{Hartmann94} applied this magnetospheric accretion model to the formation of emission lines in the spectrum of T~Tauri stars.
This fundamental paper sets the general theoretical framework for the density and temperature distributions in aligned axisymmetric magnetospheres. 
The coupling between this representation of magnetospheric accretion in T~Tauri systems with radiative transfer calculations has provided a crucial tool to interpret spectroscopic, photometric, and interferometric observations.
The sensitivity of hydrogen lines to the parameters of the magnetospheric models was studied in detail by \cite{Muzerolle2001}, improving the earlier calculations by \cite{Hartmann94}.

Near-infrared observations of the \Brg{} (\brg{}) line with the Very Large Telescope Interferometer (VLTI) GRAVITY instrument \citep{2017A&A...602A..94G} also probe the inner part of the star-disc interaction region \citep{GarciaLopez20,Bouvier20}. However, it is still difficult to associate the characteristic sizes derived from interferometry with the actual size of the magnetospheric accretion region, a key parameter in our understanding of the star-disc interaction.

In this paper, we aim at studying the formation of the \brg{} line and compute its spectroscopic and interferometric signatures for non-axisymmetric models of the inner star-disc interaction region, akin to state-of-art MHD simulations \citep{Romanova16_book}. In particular, we want to clarify the meaning of the sizes inferred through near-infrared interferometric observations and how they compare with the overall size of the magnetospheric accretion region.

In sections \S \ref{sect:rt_framework} and  \S \ref{sect:MA_flow}, we describe the model used to compute the line formation in accreting magnetospheres. We discuss spectroscopic and interferometric signatures in sections \S \ref{sect:spectro} and \S \ref{sect:interfero}, respectively.


\section{Radiative transfer framework}\label{sect:rt_framework}

We use the code MCFOST \footnote{\url{https://github.com/cpinte/mcfost} }\citep{Pinte2006,Pinte2009,tessore21} to compute emergent line fluxes from multidimensional models of magnetospheres for a 20-level hydrogen atom. The atomic model, with 19 bound levels and the ground state of \textsc{H}II, consists of 171 bound-bound transitions (atomic lines) and 19 bound-free transitions (continua). 
We focus here on the \brg{} line at 2.1661 µm although, the Balmer lines \ha{} and \hb{} and the \Pab{} line (\pab{}) are modelled as well.
These specific hydrogen lines are commonly used to characterise accretion and ejection phenomena in young systems \citep{FolhaEmerson01, Alencar2012, Bouvier20, Pouilly20, Sousa21}. The method to solve for the non-LTE populations of hydrogen and the microphysics are the same as in \cite{tessore21}. The updated version of the code we use now simultaneously solves the charge equation and the statistical equilibrium equations, which has been proven to increase the convergence in chromospheric conditions \citep{2007A&A...473..625L}.
We tested our code for different magnetospheric models taken as benchmarks in \cite{Muzerolle2001} and \cite{kurosawa06}. The results of this comparison are presented and discussed in Appendix \ref{app:benchmark}.

\section{Magnetospheric accretion model}\label{sect:MA_flow}

Matter from the circumstellar disc is channelled onto the stellar surface along the dipolar magnetic field lines. The stellar magnetic field truncates the disc at a distance $R_{t}$ from the star, the truncation radius. In practice, the interaction between the stellar magnetic field and the disc takes place over a small region between $R_t$ and $R_t + \delta r$. Both $R_{t}$ and $\delta r$ are used to define the size of the disc region magnetically connected to the star.
As the gas approaches the stellar surface, it decelerates in a shock and is heated at coronal temperatures. Theoretical models of accretion shocks by \cite{CalvetGullbring98} show that the optically thin emission of the pre/post-shock dominates below the Balmer jump and that the optically thick emission of the heated photosphere contributes to the total continuum emission at larger wavelengths. In the following, we only consider the contribution of the heated photosphere to the shock radiation. The shock\footnote{We assume that the shock region is unresolved and is part of the stellar surface.} temperature is computed from the energy of the gas infalling onto the stellar surface following the prescription of \cite{2004ApJ...610..920R} unless specified. This approach assumes energy conservation and that the shock radiates as a black body, meaning that its temperature is determined by the specific kinetic energy and enthalpy of the gas deposited at the stellar surface. The shock temperature hence derived is of the order of 4500 K - 6000 K.

\subsection{The stellar surface}
The stellar surface is considered as the inner boundary of the model and emits as a blackbody whose temperature is determined by the stellar parameters. Throughout the paper, the stellar parameters are $T_{\ast} = 4,000$ K, $M_{\ast} = 0.5$~\Msun{}, and $R_{\ast} = 2$~\Rsun{}. We set the distance to the star at 140 pc, which is typical of the nearest star forming regions such as Upper Scorpius \citep[$\approx$~146 $pc$][]{Galli18_upsco} or Taurus \citep[$\approx$~130 $pc$][]{Galli18_taurus}.

\subsection{Geometry of the accretion funnels}
We consider 3D non-axisymmetric models of the magnetospheric accretion region.
These models are parametrised by the same set of parameters as the axisymmetric magnetospheric model of \cite{Hartmann94} \citep[see also][]{Muzerolle98,Muzerolle2001,kurosawa06,Lima2010,Kurosawa2011,Dmitriev19}.
The density and the velocity fields of the accretion columns are fully described with a set of independent parameters: the mass accretion rate $\dot{M}$, the rotation period $P_{rot}$, $R_t$, and $\delta_r$.

\begin{figure}
    \centering
    \includegraphics[width=0.9\columnwidth]{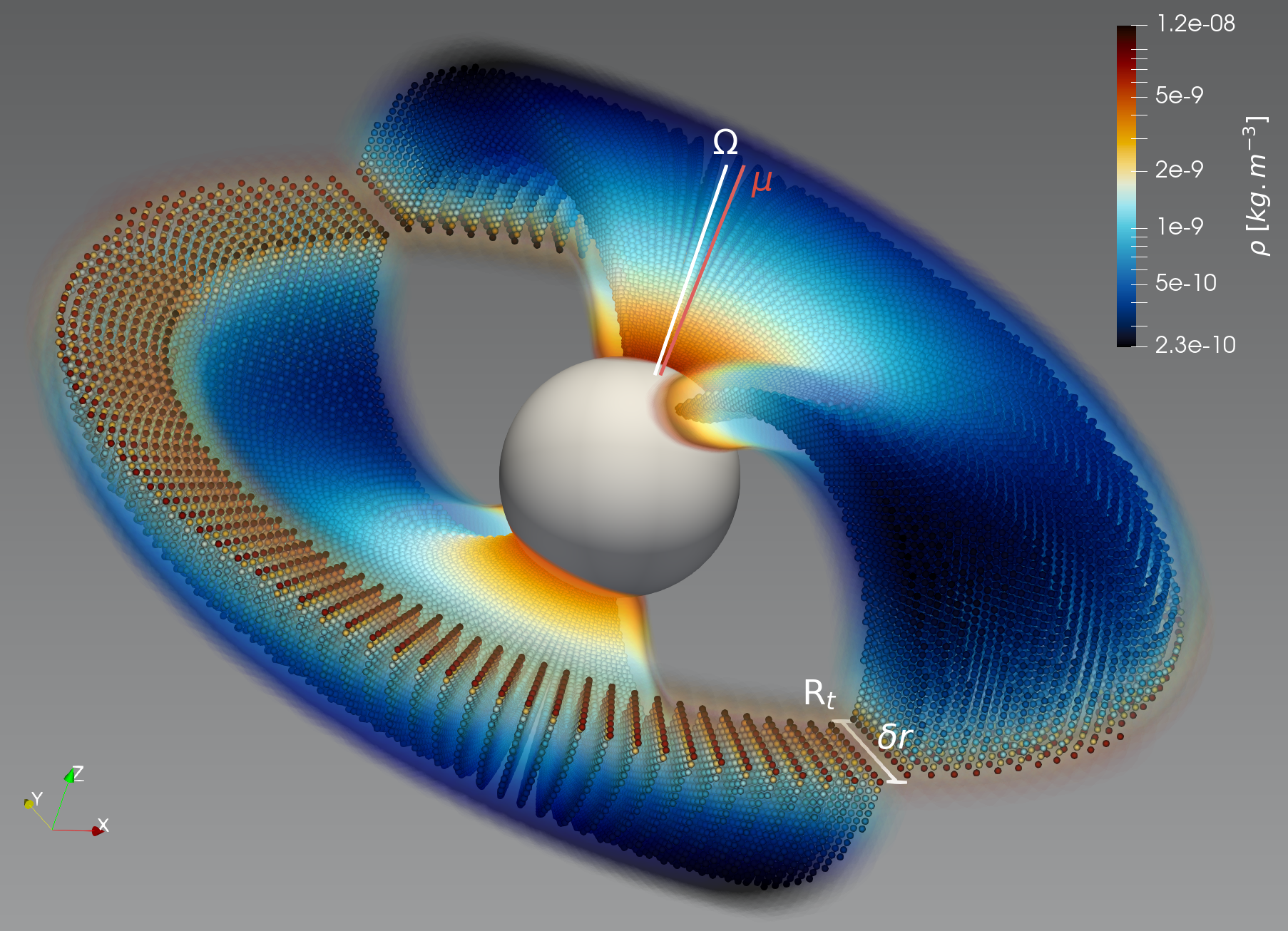}
     \caption{Density distribution of a non-axisymmetric model with an obliquity of 10$^\circ$. The rotation axis of the star $\Omega$ is shown with a white arrow and the dipole axis, $\mu$, with a red arrow.
     The density is computed from Eqs.~(\ref{eq:ideal_mhdeq}) and (\ref{eq:mass_acc}). The colour map scales with the density.}
     \label{fig:density_obliquity_10}
\end{figure}

For our study, the value of $\dot{M}$, $R_t$ and $\delta_r$, and of the temperature of the magnetosphere are fixed. The impact of these parameters on the line formation has been discussed thoroughly in \citet[][see also App. \ref{app:benchmark}]{Muzerolle98,Muzerolle2001}. 
The line's response to the mass accretion rate and to the temperature is an essential proxy for understanding the physics of the star-disc interaction region.
We use a mass accretion rate $\dot{M} = 10^{-8}$~\Msunyr{}, a truncation radius $R_t = 4\, R_{\ast}$, and $\delta_r = 1 \, R_{\ast}$. 
The value of the rotation period is deduced from the maximum truncation radius ($R_t + \delta_r$), imposing that stable accretion occurs at 90\% of the corotation radius, consistent with the work of \cite{Blinova16}. The rotation period is therefore fixed at $P_{rot} = 6$ days, corresponding to slowly rotating T~Tauri stars \citep[see][for a review]{HerbstPP07,BouvierPP14}. The rotational velocity for that period is thus of the order of 80~\kms{} at the outer edge of the magnetosphere. 

When the magnetic field axis ($\mu$) is misaligned with respect to the rotational axis ($\Omega$) of the star, the geometry of the accretion flow changes dramatically. The equations for the magnetic field components of a non-axisymmetric dipole, i.e. with a non-zero obliquity, are provided in \cite{Mahdavi&Kenyon98}. The parameter $\beta_{ma}$ describes the angle between the dipole moment and the star's rotational axis, the magnetic obliquity.

We approximate the density, $\rho$, along the non-axisymmetric magnetic field lines with,
\begin{equation}\label{eq:ideal_mhdeq}
    \rho = \alpha \dfrac{B}{\varv}= 
    \alpha B\dfrac{\rho_{axi}}{B_{axi}} \text{,}
\end{equation}
where $\alpha$ is a constant along a given field line and $B$ the analytic misaligned dipolar field. $\varv$, $\rho_{axi}$, and $B_{axi}$ denote the velocity field, density, and dipolar magnetic field, respectively, and they are taken from the axisymmetric model of \citet{Hartmann94}. In other words, the 3D density structure is computed from Eq.~(\ref{eq:ideal_mhdeq}) under the assumption that the infalling gas has a velocity field on the poloidal plane.
The value of $\alpha$ is computed from the numerical integration \footnote{For this 3D magnetospheric accretion model, an explicit formula for the shock area does not exist \citep[see also][]{Mahdavi&Kenyon98}} of the mass flux over the shock area, 
\begin{equation}\label{eq:mass_acc}
    \dot{M} = \displaystyle\int \rho \mathbf{v} \cdot \mathbf{dS} \text{,}
\end{equation}
where $dS$ is the surface element and $\mathbf{v}$ the velocity field.
In our model, the value of $\dot{M}$ is an input parameter and is held constant. Therefore, $\alpha$ is obtained to ensure consistency between Eqs.~(\ref{eq:ideal_mhdeq}) and (\ref{eq:mass_acc}). 
We compute five models with an obliquity $\beta_{ma}$ ranging from \textbf{five} to forty degrees in step of ten degrees, representative of what has been measured for T~Tauri stars with spectroscopy \citep{McGinnis20} and spectropolarimetry \citep{Donati08,Donati10,Donati13,Johnstone14,Pouilly20}.
For these non-axisymmetric models, the shortest field lines -- defining the main accretion columns\footnote{Geometrically, the shortest field lines obey the following criterion $\cos \phi^{\prime} \times z >0 $ where $\phi^{\prime}$ is the azimuth in the frame aligned with the dipole axis and $z$ the coordinate parallel to the rotation axis.} -- carry most of the gas density. We remove the longest field lines -- the secondary columns -- in our modelling as in \cite{Esau2014}. This yields models with one crescent-shaped accretion spot per stellar hemisphere reminiscent of numerical simulations of misaligned dipoles \citep{Romanova03}.

Figure \ref{fig:density_obliquity_10} shows the density of a non-axisymmetric magnetosphere with an obliquity of 10$^{\circ}$.

\subsection{Temperature of the funnels}
The temperature of the magnetospheric accretion region is not well constrained. The determination of the temperature by \cite{Martin96} from first principles was not able to reproduce the observations. A self-consistent calculation of the temperature of the magnetosphere is beyond the scope of this paper. Instead, we adopt here the temperature profile of \cite{Hartmann94}, which has been extensively used in the past to model line fluxes from accreting T~Tauri stars.
The temperature is computed using a volumetric heating rate ($\propto r^{-3}$) and balancing the energy input with the radiative cooling rates of \cite{Hartmann82}. 
The exact balance between the heating and cooling mechanisms is unknown. Instead, the temperature profile is normalised to a free parameter, $T_{max}$, that sets the value of the maximum temperature in the funnel flow.
In the following, we have set the temperature maximum to $T_{max} = 8,000$ K.

\section{Spectroscopic signatures}\label{sect:spectro}
\begin{figure*}[h!]
    \centering
    \includegraphics[width=1\textwidth]{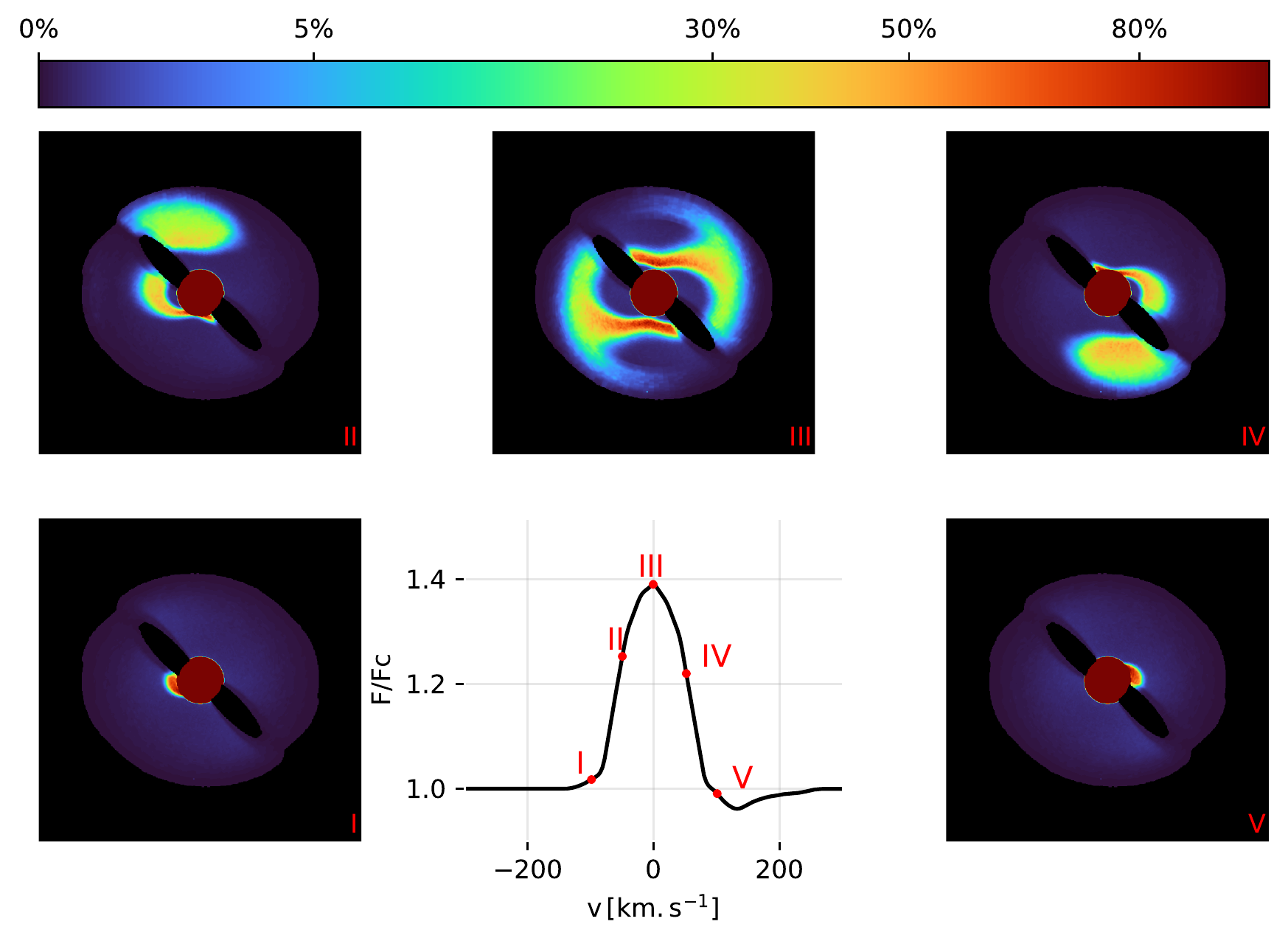}
     \caption{Origin of the emission seen across the \Brg{} line. The contribution of individual images to the total line flux is indicated on the central image showing the line profile. The brightness maps are in units of the maximum emission. The emission of the stellar surface is saturated. Orange to red colours indicates the regions of maximum emission. The system is seen at an inclination of 30$^{\circ}$ and an rotational phase of $\sim$0.25, similar to Fig. \ref{fig:density_obliquity_10}.}
    \label{fig:line_mosaic_Brg}
\end{figure*}

Thanks to the Doppler shift of the funnel flow, it is possible to reconstruct the origin of the emission line by looking at the brightness maps in various velocity channels. Figure \ref{fig:line_mosaic_Brg} shows the contribution of the different parts of the magnetosphere to the total integrated \brg{} line flux at a given velocity for an inclination of 30${}^{\circ}$, matching the model illustrated in Fig. \ref{fig:density_obliquity_10}.
At those density and temperature, the continuum emission comes from the stellar surface (${\rm I_{surf} / I_{mag} > 100 }$). Locally, the continuum emission from the shock is three times larger than the emission from the star. Overall, given the small covering area of the accretion shock (around 1\%), the total continuum emission at the frequency of the \brg{} line is dominated by the star's radiation, ${\rm F_{shock}/F_{\ast} = 3\%}$.
The low-velocity components (< 50~\kms{}) of the line form in the regions where the projected velocity along the line-of-sight is close to zero and near the disc.
The geometry of the non-axisymmetric model, defined in \S \ref{sect:MA_flow}, is responsible for a rotational modulation of the integrated line flux. 
Many classical T~Tauri stars shows modulated photometric variability \citep[e.g][]{Cody14} and, more directly related to the magnetospheric region, many also show rotational modulation of the longitudinal component of the stellar magnetic field \citep[e.g][]{Donati20}. Indeed, the periodic variability of optical and emission line profiles has been reported in various systems \citep[for instance][]{sousa16,Alencar18,Bouvier20}, which indicates that the emission region is stable on a timescale of several rotation periods.

\begin{figure*}[h!]

  \includegraphics[width=1\textwidth]{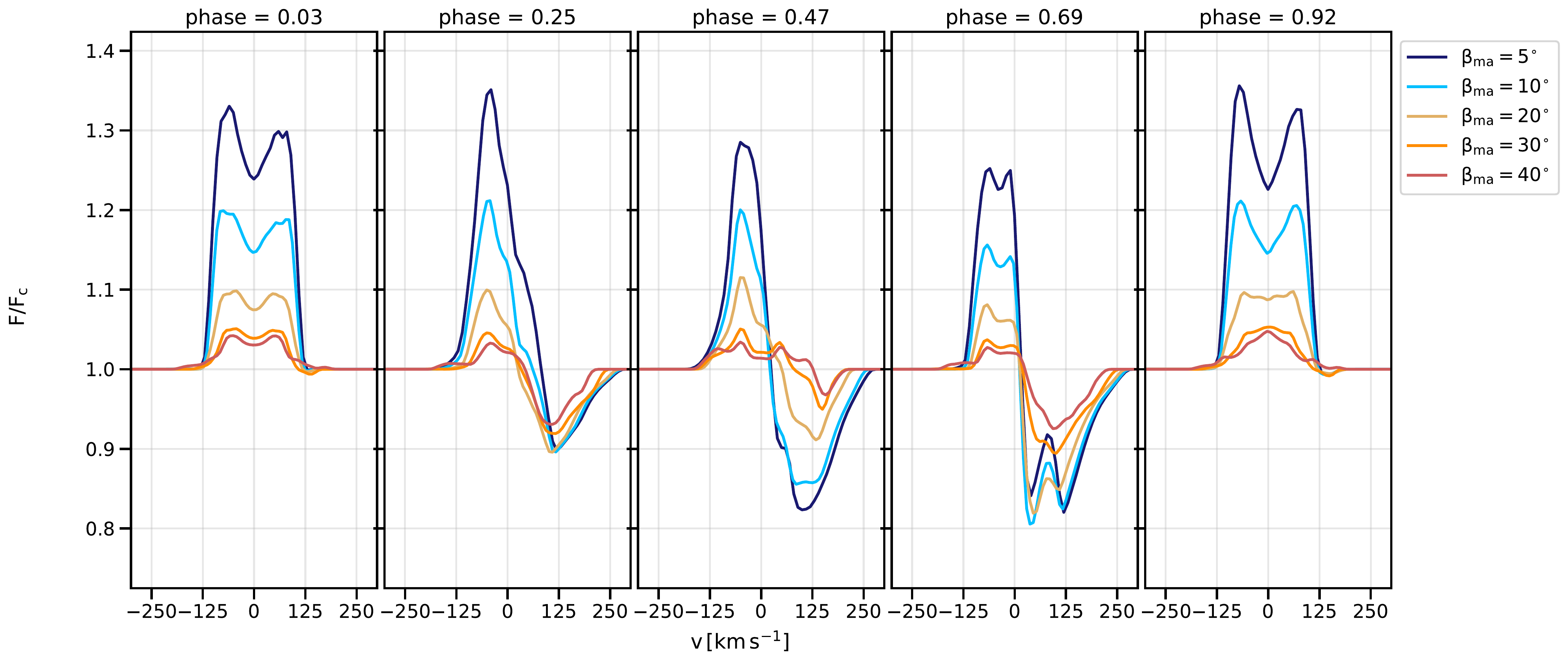}
 \caption{\Brg{} line variability along the rotational cycle. Each column corresponds to a specific rotational phase. At phase 0, the shock area is unseen on the stellar surface, while phase of 0.5, the shock is fully seen on the visible hemisphere. The colours correspond to different values of the obliquity. All fluxes are computed with an inclination of 60$^{\circ}$.}
     \label{fig:brg_allazim}
 \end{figure*}

Figure \ref{fig:brg_allazim} shows the variability of the \brg{} line at different phases of rotation at an inclination of $60^{\circ}$ for different obliquities.
The origin of the rotational phase is defined such that at phase 0.5, the accretion shock is facing the observer.
The red-shifted absorption seen for the \brg{} line at phases 0.250, 0.47 and 0.69, results from a lower source function of the gas above the shock (see App. \ref{app:benchmark}). From observations, red-shifted absorption in the \pab{} and \brg{} lines are seen in less than 34\% and 20\% of the line profiles, respectively \citep{FolhaEmerson01}. The inverse P~Cygni profile disappears when the shock, or a significant fraction of it, is hidden on the opposite side of the star. The line, with either a double-peaked profile or a moderate red-shifted absorption, is reminiscent of \cite{Reipurth96} cases II and IV.
While the profiles with redshifted absorption agree with observations, those that display an M-shape are usually not observed in young stellar objects. This suggests that magnetospheric accretion is not the only contribution to the profile, which can also be impacted by various types of outflows \citep[e.g., stellar, interface, and disk winds][]{Lima2010,Kurosawa2011}. The optically thick accretion disc is not included in our models. The effect of the disc emission and absorption on the spectroscopic and interferometric observables will be discussed in a subsequent paper.

\begin{figure}[H]

    \includegraphics[width=0.98\columnwidth]{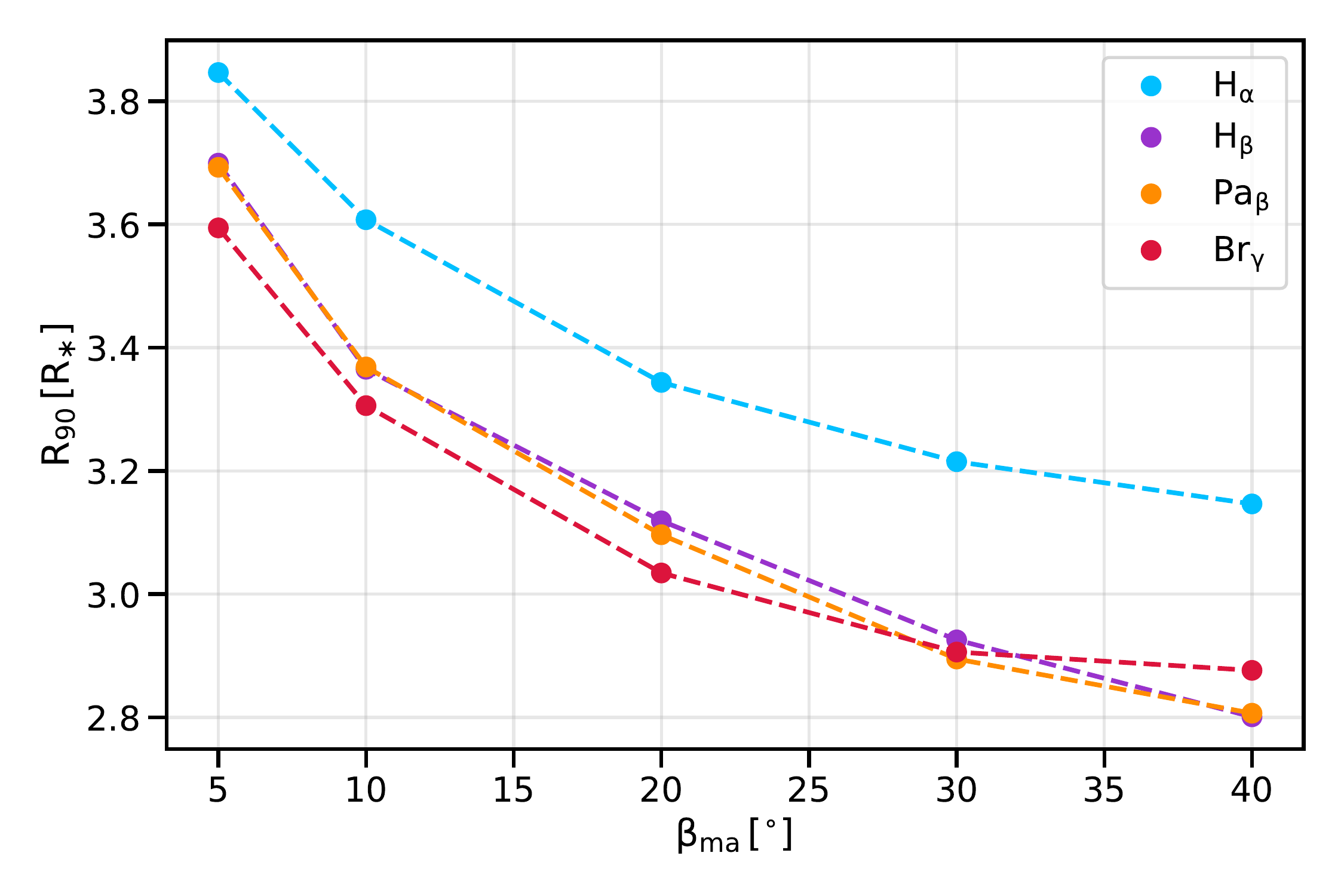}

    \caption{Radius encompassing 90\% of the total flux ($R_{90}$) for each line as a function of the obliquity, $\beta_{ma}$. Hydrogen lines are labelled with different colours.}

    \label{fig:radius_90}
\end{figure}

We also observe a decrease of the line flux as the obliquity increases. Figure \ref{fig:radius_90} shows the radius encompassing 90\% of the total line flux, $R_{90}$, at an inclination of 60$^{\circ}$ for non-axisymmetric models with different obliquities for the \ha{}, \hb{}, \pab{} and \brg{} lines.

As $\beta_{ma}$ increases, the volume of the magnetospheric accretion region decreases because the arc length of the  accreting field lines shortens. Therefore, the total flux, for all lines, decreases accordingly, independently of the viewing angle of the system. However, we also note a dependence of $R_{90}$ with the line. The value of $R_{90}$ represents the size of the emitting region in  a given line, which is a function of density and temperature, and of the viewing angle.

\section{Interferometric signatures}
\label{sect:interfero}

In this section, we compute the size of the \brg{} line-emitting region inferred from interferometric observations, and we compare it to model flux radii (see \S\ref{sect:spectro}).

\subsection{Interferometric observables}

The interferometric observables are derived from the radiative transfer (RT) model using the \texttt{ASPRO2}\footnote{Available at \url{https://www.jmmc.fr}} software developed by the Jean-Marie Mariotti Center (\texttt{JMMC}). These observables represent what would be observed with GRAVITY in the near-infrared. We consider the configuration obtained with the Very Large Telescope (i.e. 4x8m telescopes), encompassing a range of baselines from 35 to 135 m. With a typical night of 8 hours, we compute one observing point per hour for the six baselines of the VLTI to increase the Fourier sampling, namely u-v coverage, which is crucial for the fitting part of our approach. As described in \citet{bourges16}, we derive the observables from the RT images (see Fig. \ref{fig:line_mosaic_Brg}) by computing the complex visibility in each spectral channel around the \brg{} line and interpolating them to match GRAVITY's spectral resolution (R = 4000). Specifically, we simulate a total of 37 spectral channels (from 2.161 to 2.171 µm with a step of $2.8 \,10^{-4}$ µm) for the six projected baselines repeated eight times. Within this range, 31 spectral channels are used to measure the K-band continuum and six channels sample the \brg{} line emitting region.

Figure \ref{fig:uv_coverage} illustrates the resulting u-v plane projected on-sky for a typical object observed at the VLTI with a declination of -34$^\circ$ (e.g. TW Hydrae).

Figure \ref{fig:interfero} shows the interferometric observables along the rotational cycle for a model with an inclination of 60$^{\circ}$ and an obliquity of 10$^{\circ}$. The two main observables are: the modulus of the complex visibility -- the visibility amplitude -- and the differential phase -- its argument -- dispersed in wavelength. The phase is normalised to zero in the continuum. The visibility amplitude can then be used to estimate the object's size, while the phase measures the photo-centre shifts between the line-emitting region and the continuum. The phase can only be used as a relative measurement (e.g. between the line and the continuum), the absolute phase being lost due to a combination of atmospheric and instrumental effects. We repeat the simulated observations and compute nine datasets over a rotational cycle sampled every 40 degrees (~0.11 in phase). 
In this study, we are interested in the line's emitting region only. Therefore, we use pure line quantities, instead of total visibilities and phases, to remove the contribution from the stellar surface (see appendix \ref{app:interfero} for the derivation of the pure line interferometric quantities).

\begin{sidewaysfigure*}[h!]
    \includegraphics[width=\textwidth]{fig/interfero/resum_results_Bgr_tilt=10_v4.pdf}
    \caption{Synthetic interferometric measurements and modelling across the rotational phase of the system. \textbf{Top:} integrated images over the \brg{} line. Green (uniform disc) and yellow (Gaussian disc) ellipses are the characteristic sizes measured with a GRAVITY-like instrument. \textbf{Middle-top:} Pure line visibility amplitude observables associated with the corresponding models. The visibility variation (for a given baseline) as the u-v plane rotates is the specific signature of an elongated object. \textbf{Middle-bottom:} Pure phase visibility across the line profile for the six baselines of the VLTI. Colours encode the observing time. \textbf{Bottom:} Velocity map of the radiative transfer model. The coloured dots represent the measurement of the photo-centre derived from the phase visibility in each available spectral channel (see appendix \ref{app:interfero}). In each figure, the magnetosphere and the stellar surface have been normalised independently, for display purpose.}
    \label{fig:interfero}
\end{sidewaysfigure*}

\subsection{Physical characteristic and sizes}

Once the interferometric observables are computed, we apply standard modelling methods to interpret the data \citep{2003EAS.....6...23B}. Firstly, we average the visibility amplitude of the six spectral channels available within the \brg{} line\footnote{Five and four spectral channels only were used at phase 0.25 and 0.47, respectively, due to a limited line-to-continuum ratio (see Appendix \ref{app:interfero} for details).}. We use the average visibilities to recover the global size of the \brg{} emitting region, where the different velocities probe specific parts of the moving material within the magnetosphere. Then, we fit the averaged visibility amplitude using elongated Gaussian or uniform disc models. Such models are typically used in interferometry to estimate the system's characteristic size and on-sky orientation. The source's brightness distribution is defined by its half-flux radius in the case of a Gaussian disc or its radius for the uniform disc model, and an elongation factor and a position angle. In the following, we adopt the definition of "radius" for both models, which corresponds to the half-flux semi-major axis for the Gaussian model and the semi-major axis for the uniform disc model. The recovered sizes and orientations are represented in the top panel of Fig. \ref{fig:interfero}. While neither model can fully account for the size of the magnetosphere, the uniform disc probes a larger area of the magnetosphere, while the Gaussian disc seems limited to the most luminous parts. We note that the fit of the visibility is equally good for both models and, thus, does not allow us to discriminate between the models from the synthetic visibilities only (middle-top, Fig. \ref{fig:interfero}). 

\begin{figure}[H]
    \includegraphics[width=0.99\columnwidth]{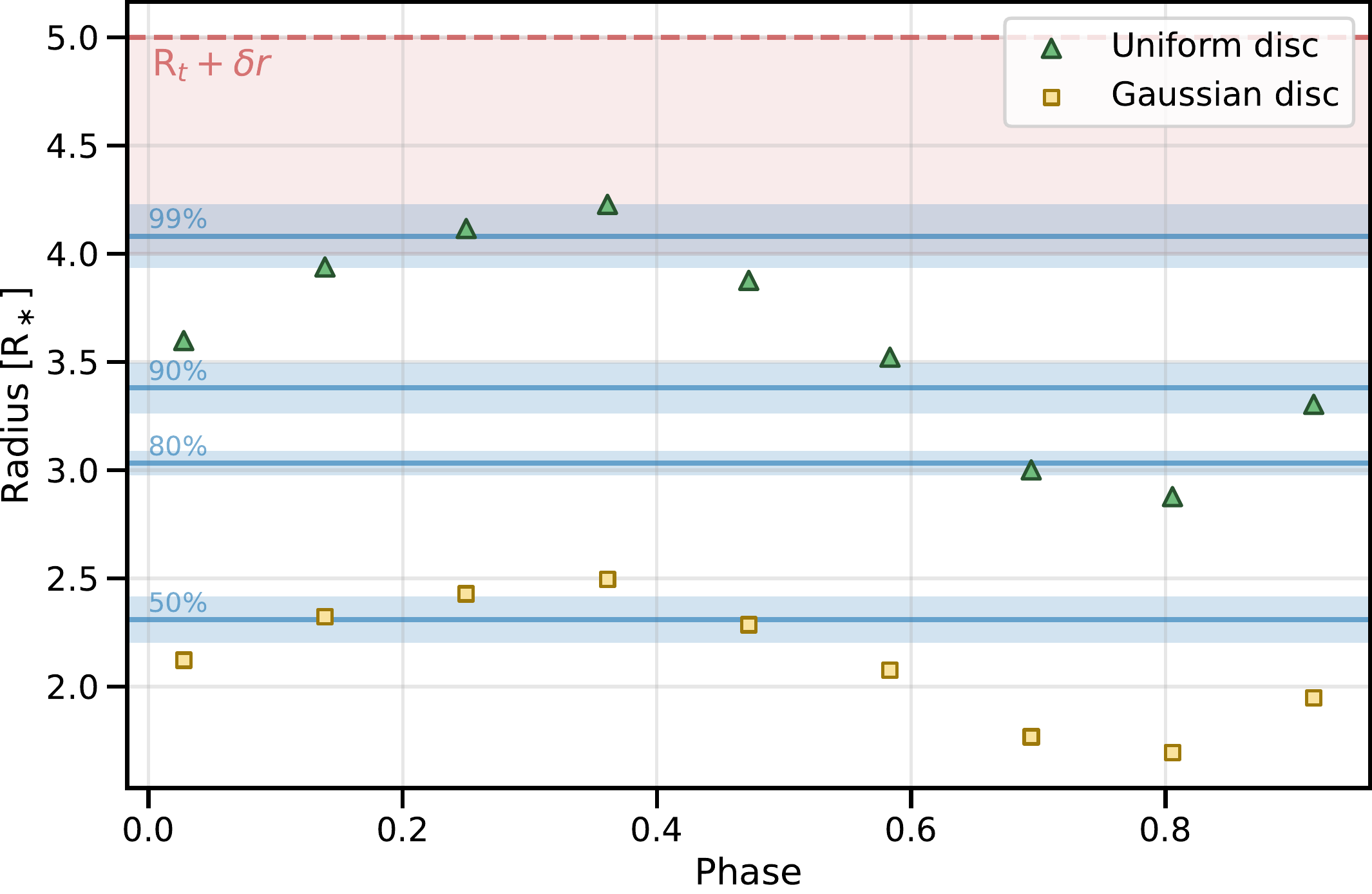}
    \caption{Interferometric radii as a function of the rotational phase. Uniform and Gaussian disc models are shown with green and yellow markers, respectively. Blue lines correspond to the radii encompassing 50, 80, 90 and 99$\%$ of the total RT model's flux. The blue shaded areas represent the standard deviation of these radii across the rotational phase.} The red shaded area indicates the inner ($R_t$) and outer radius ($R_t + \delta r$) of the RT model.
    \label{fig:size_azimuth}
\end{figure}

In order to quantify the physical meaning of the interferometric measurements, we compare the interferometric sizes with reference flux radii of the RT models. We set these radii to represent 50, 80, 90, and 99$\%$ of the total flux emitted by the magnetospheric accretion region. Figure \ref{fig:size_azimuth} compares the sizes derived with interferometry to the characteristic radii of the RT models.

We find that the size derived from the uniform disc model is modulated around an average value of 3.5 R$_\ast$ corresponding to 90$\%$ of the \brg{} emitting region. The size obtained by interferometry appears to be modulated by the position of the funnel flows close to the star, with a minimum located around phase 0.8. The Gaussian model exhibits the same modulation but with a lower amplitude ($2.1\pm0.4$ R$_\ast$) and appears sensitive to the magnetosphere's innermost region, close to the 50$\%$ flux radius. The size derived from the uniform disc model emerges as being the most appropriate to recover the reference model size, accounting for at least 80$\%$ of the total flux emitted by the magnetosphere. 

The derived orientations obtained from interferometry seem to be particularly representative of the position of the accretion funnel flow and the on-sky orientation of the \brg{} emitting region (Fig. \ref{fig:interfero}). The measured position angle agrees with the magnetosphere's orientation, particularly when the shock faces the observer (phase = 0.5). Nevertheless, it appears somewhat hazardous to decipher the shape and orientation of the emitting region across the rotational cycle from this observable only, as different magnetospheric configurations can be described by a very similar interferometric model (e.g. phases 0.03 and 0.25). A stronger constraint on the orientation of the funnel flows arises from  differential phase measurements.

\subsection{Differential phases and photo-centre shifts}
From the differential phases, we can derive the photo-centre shift between the continuum and the \brg{} line emitting region. In the regime of marginally resolved sources, there is a direct relationship between the projected photo-centre displacement vector (\textbf{P}) and the phase along each baseline \citep{Lachaume03}:
\begin{equation}
    \phi_i = -2\pi\frac{B_i}{\lambda}\textbf{P},
\end{equation}
where $\phi_i$ is the differential phase measured for the $i$th baseline, $B_i$ is the length of the corresponding baseline, and $\lambda$ is the effective wavelength of the spectral channel. A four telescope beam-combiner like GRAVITY gives access to six projected baselines that enable us to accurately retrieve the value and orientation of the photo-centre shifts in each spectral channel \citep{LeBouquin09, Waisberg17}. Such a measurement results in a position-velocity plot of the displacement of the photo-centre across the \brg{} line relative to the continuum. This is illustrated in the bottom panels of Figure \ref{fig:interfero}.

The photo-centre shifts trace the accretion funnel flow's direction and follow the stellar rotation. For instance, when the northern accretion shock (N-shock) is located behind the star (phase = 0), the accreting material falls onto the stellar surface in the direction of the observer. Accordingly, the photo-centre measured in the blue-shifted part of the line profile ($\simeq$ -75~\kms{}) lies on the blue-shifted part of the velocity map, corresponding to the approaching funnel flow. Equivalently, the photo-centre measured in positive velocity channels of the line profile ($\simeq$ +75~\kms{}) is shifted towards the receding funnel flow. In contrast, when the shock faces the observer (phase = 0.5), the velocity map goes from blue to red in the east-west direction, and the photo-centre shifts recover this trend as demonstrated at phase 0.47. 

We can thus identify three privileged directions and shapes of the photo-centre shifts: -- linear north-south at phase $\simeq0$ (N-shock behind), -- S-shape at phase 0.25 and 0.69 and -- linear east-west at phase $\simeq0.5$ (N-shock in front). The differential phase is, therefore, a key ingredient to recover the geometry and orientation of the line-emitting region, tracing the moving material along a rotational cycle.

\subsection{Signal-to-noise considerations}

As a proof-of-concept, the results presented above assume infinite signal-to-noise ratio. The goal is to predict the spectroscopic and interferometric signatures of the magnetospheric accretion process. Thus, the models predict  typical visibility amplitudes ranging from 1 down to 0.97 (see Fig.~\ref{fig:interfero}). Such a modest interferometric signal requires a measurement accuracy of about 1$\%$ to be securely detected. Similarly, the models predict a deviation of the differential phases by  1 to 2 degrees (Fig.~\ref{fig:interfero}), which requires an accuracy of order of a fraction of a degree to yield a robust detection. Recent interferometric studies performed with VLTI/GRAVITY in the K-band demonstrate that these levels of accuracy can be routinely obtained indeed with reasonable exposure times  on young stellar objects \citep[e.g.][]{Bouvier20a, GarciaLopez20, Wojtczak22}, or active galactic nuclei \citep{Gravitycollab_sturm18}.

\section{Summary and conclusion}\label{sect:discussion}

We presented non-LTE radiative transfer modelling of the \Brg{} line emission for non-axisymmetric models of accreting magnetospheres.
We used the equations of a misaligned dipolar magnetic field to derive the geometry of the magnetospheric accretion region for different obliquities of the magnetic dipole.
We used MCFOST to compute radiative signatures of the \brg{} line along a full stellar rotational cycle. Further, we derived near-infrared interferometric observables for the line, comparable to what the GRAVITY instrument has already measured for T~Tauri stars.

The main conclusions of this study are the following:
\begin{itemize}
    \item[1)] The total flux in the line, and the line-to-continuum ratio, depends on the obliquity of the dipole. As the obliquity increases, the size of the emitting region decreases, leading to a lower integrated flux. Also, projection effects make the emission region of lines forming close to the stellar surface appearing narrower.
    
    \item[2)] The \brg{} line total flux varies with the rotational phase due to the non-axisymmetry of the models induced by the magnetic obliquity. The line profiles exhibit a red-shifted absorption, that is an inverse P~Cygni profile, when a significant fraction of the accretion shock is aligned with the observer's line of sight. When the shock is hidden on the opposite side of the star, the line profiles exhibit a double-peaked shape, reminiscent of the lines formed in rotating envelope. The latter is due to the relatively large rotational velocity of the magnetospheric model ($\sim$80~\kms{}).
    
    \item[3)] Near-infrared interferometric observations in the \brg{} line directly probe the size of the magnetospheric accretion region. The Gaussian disc model is sensitive to the brightest parts of the magnetosphere, up to 50\% of the truncation radius, while a uniform disc model grasps  90$\%$ of the magnetosphere. It is of prime importance to consider this aspect when estimating magnetospheric radius from interferometric measurements. In both cases, the measured radius varies with the rotational phase (due to the non-axisymmetry of the dipole). A robust interferometric estimate of the magnetospheric radius therefore requires monitoring the system over a full rotational cycle.

    \item[4)] The combined knowledge of the differential phase and the associated photo-centre shifts gives hints on the object orientation and geometry. More specifically, the relative direction of the photo-centre shifts indicates the changing orientation of the accreting material along the rotational cycle in the non-axisymmetric case.
\end{itemize}

Near-infrared interferometry of the \Brg{} line is used to characterise the inner star-disc interaction region, and offers a good estima   te of the size of the line's forming region, at sub-au precision. Comparing this size with reference model radii, such as the truncation radius, allows us to distinguish between multiple origins of the \brg{} line, within or beyond these radii (e.g. magnetosphere, stellar and disc winds, jets). Further, simultaneous spectroscopic and interferometric observations along a rotational cycle, have the potential to unveil the geometry and orientation of the line's emitting region.
The variability of the line associated with the photo-centre shifts, provides a unique and unambiguous proxy of the physical processes occurring in the magnetosphere of young accreting systems, within a few hundredths of an astronomical unit around the central star.

\appendix
\section{benchmark}\label{app:benchmark}
\begin{figure*}[h!]

    \includegraphics[width=\textwidth]{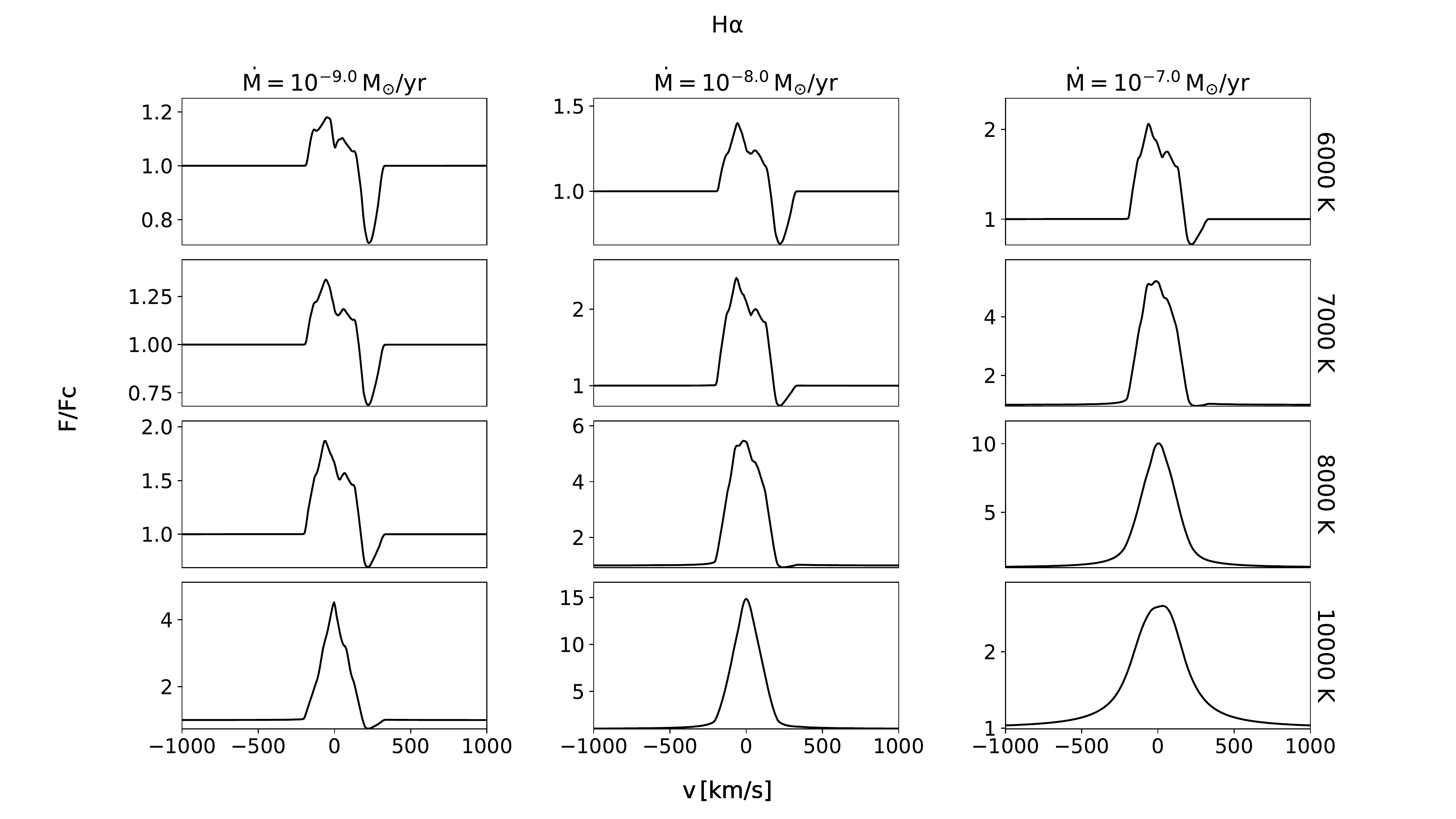}

    \caption{Dependence of \ha{} line flux with mass accretion rates $\dot{M}$ and maximum temperatures $T_{max}$. The inclination of the system is $60^{\circ}$.}

    \label{fig:figure_Ha}
 \end{figure*}

\begin{figure*}[h!]

    \includegraphics[width=\textwidth]{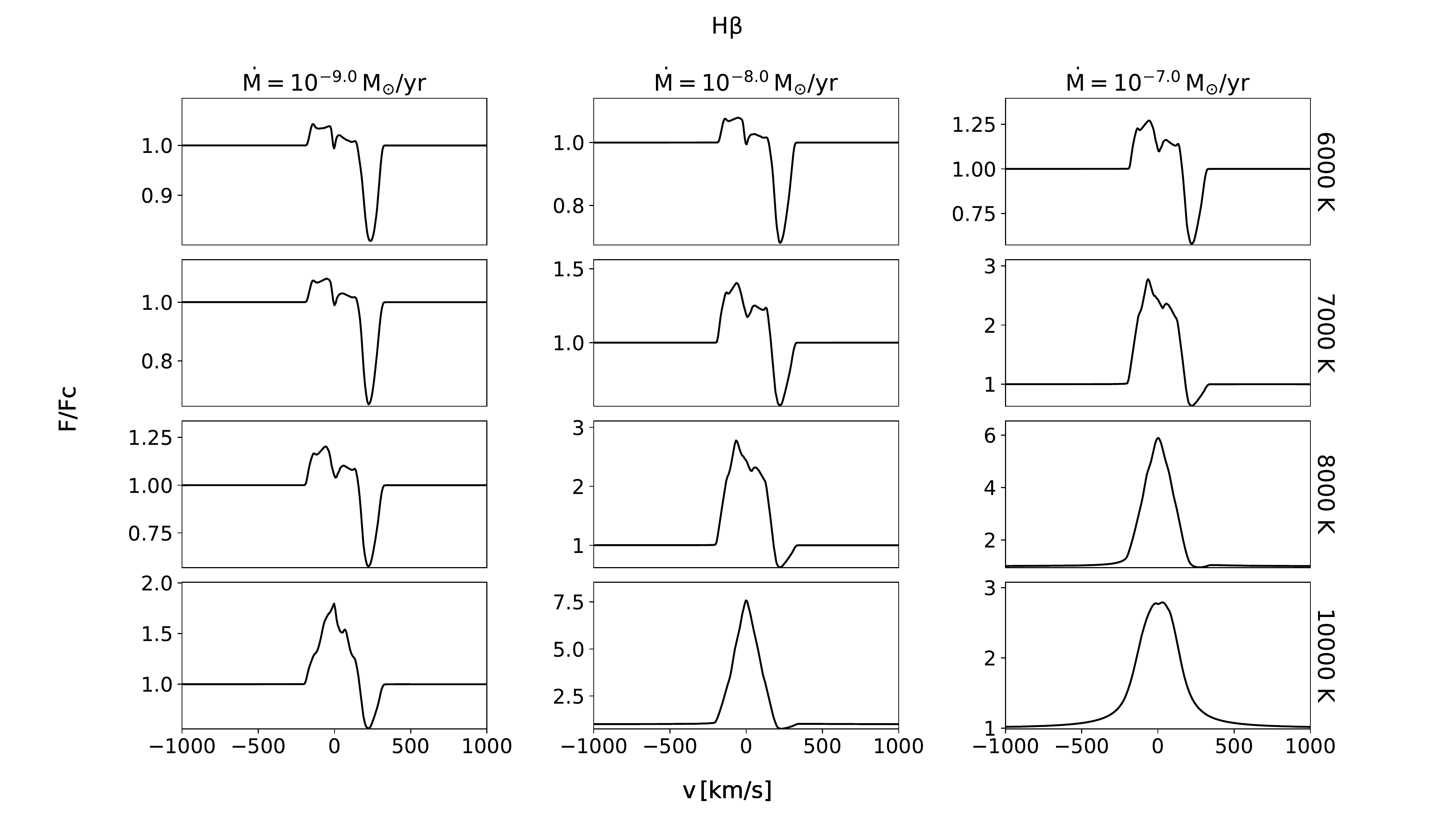}

    \caption{Same as Fig. \ref{fig:figure_Ha} for \hb}

    \label{fig:figure_Hb}
 \end{figure*}

\begin{figure*}[h!]

    \includegraphics[width=\textwidth]{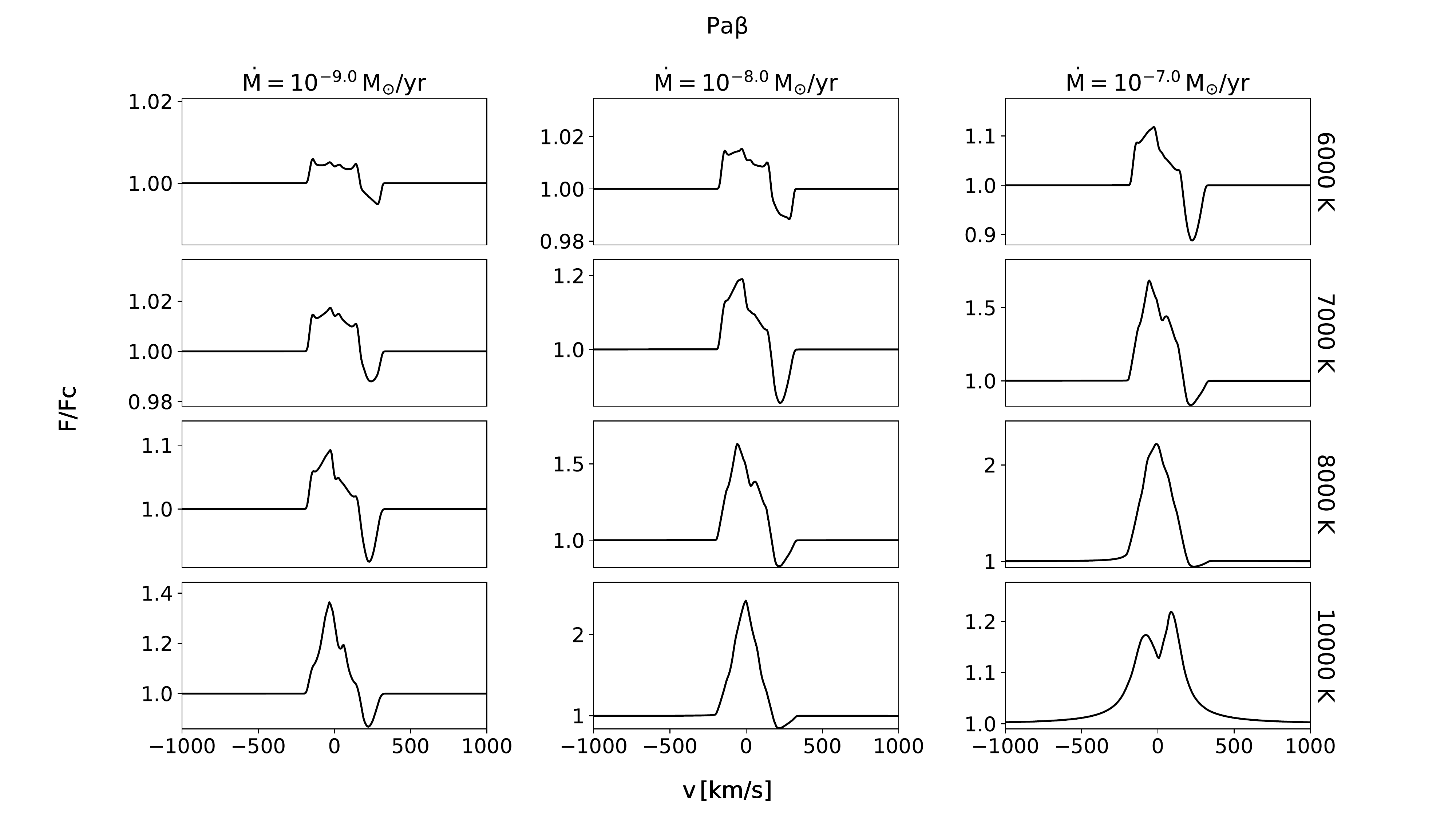}

    \caption{Same as Fig. \ref{fig:figure_Ha} for \pab.}

    \label{fig:figure_Pab}
 \end{figure*}

\begin{figure*}[h!]

    \includegraphics[width=\textwidth]{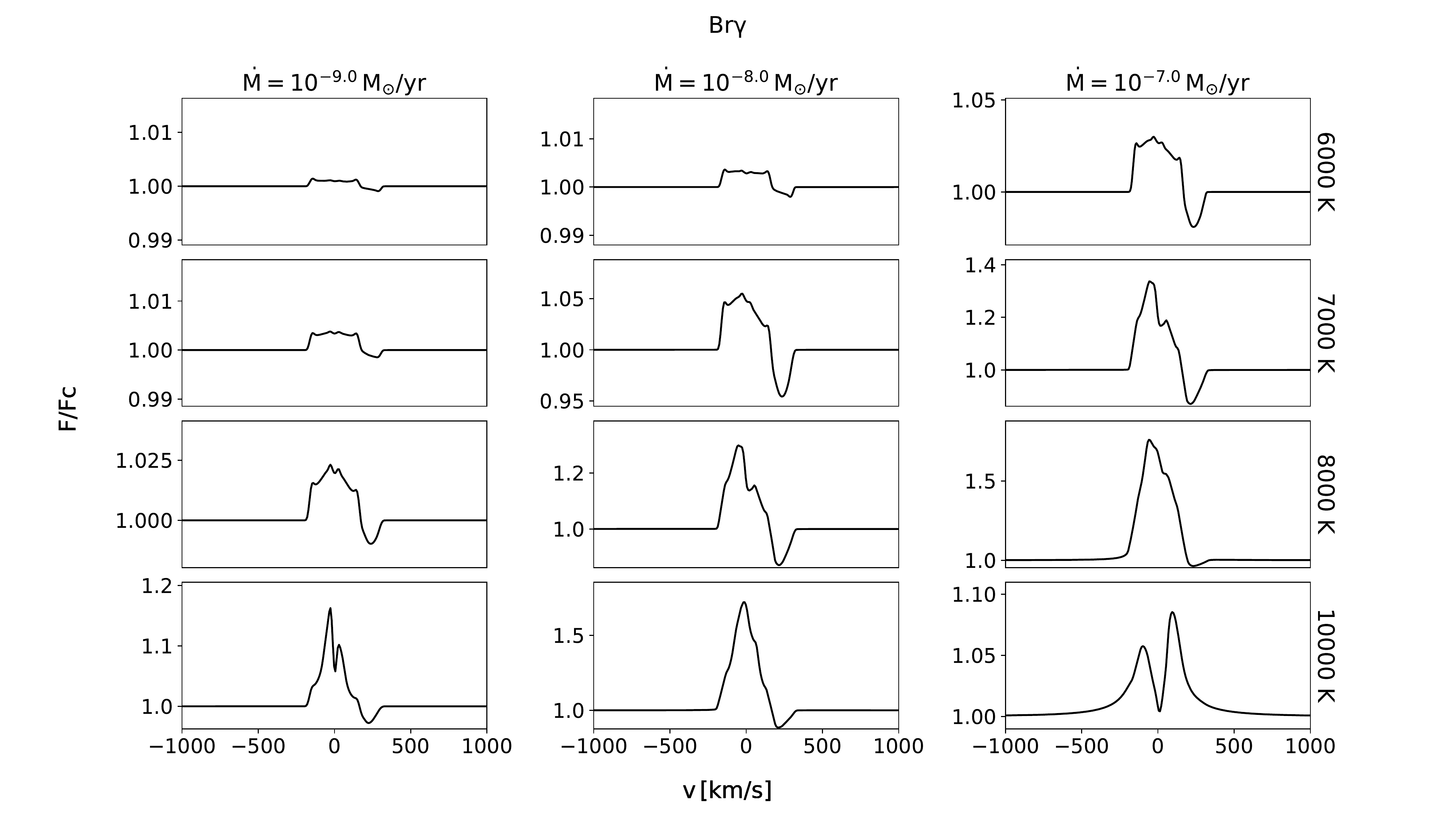}

    \caption{Same as Fig. \ref{fig:figure_Ha} for \brg.}

    \label{fig:figure_Brg}
 \end{figure*}

We present here the comparison between line profiles obtained with MCFOST and previous studies. The magnetospheric model corresponds to the axisymmetric and compact configuration of \cite{Muzerolle2001} with a fixed shock temperature at 8 000 K, a rotation period of 5 days and the following canonical T Tauri parameters: $M_{\ast}=0.8\,M_{\odot}$, $R_{\ast} = 2\, R_{\odot}$ and $T_{\ast} = 4\,000 \, K$. The inclination of the system is 60 degrees. The continuum emission of the stellar surface (shock and photosphere) is constant for all models.
Figures \ref{fig:figure_Ha}, \ref{fig:figure_Hb}, \ref{fig:figure_Pab}, and \ref{fig:figure_Brg} show the \ha, \hb, \pab{} and \brg{} lines profiles for different values of $T_{max}$ and $\dot{M}$. 
An inverse P Cygni profile, with a red-shifted absorption, is seen for all lines although it is dependent on the value of the mass accretion rate and of the maximum temperature.
For a given mass accretion rate, an increase of the maximum temperature results in a higher line emission peak and a shallower red-shifted absorption. As the temperature increases, the line source function increases, which is the cause of a higher emission above the continuum emission.
The appearance of the red-shifted absorption component is caused by absorption from the gas above the stellar surface. It is controlled  by the ratio between the source function of the line in the accretion funnel and that of the underlying continuum from the stellar surface, especially at low mass accretion rates and temperatures.
Eventually, for the highest mass accretion rate and temperature, the lines become so optically thick that the red-shifted absorption is washed out by the large wings of the line. The red-shifted absorption is more pronounced for lines forming closer to the accretion shock like the \hb{} line.
At a temperature larger than 8,000 K and a mass accretion rate above 10$^{-8}$ \Msunyr{}, the continuum emission from the magnetosphere becomes important and the line-to-continuum ratio decreases.
This effect is seen for instance in the \ha{} line (see Fig. \ref{fig:figure_Ha}).
When the mass accretion rate increases for a given temperature, the density of the magnetosphere increases. As a consequence, the line source function increases. At high temperature and high density, the background continuum  emission  of  the  magnetosphere  dominates for certain wavelengths, and absorption occurs. The latter effect is seen in the \pab{} (Fig. \ref{fig:figure_Pab}) and \brg{} (Fig. \ref{fig:figure_Brg}) lines where the strong continuum contribution at the disc surface leads to absorption at low velocities, where the lines source function is small.
These results are consistent with the previous studies and demonstrate the robustness of our code for modelling the close environment of T~Tauri stars \citep{tessore21}.

\section{Derivation of the interferometric pure-line phase and visibility}\label{app:interfero}

We focus on the magnetospheric emission probed by the \brg{} line and, therefore, aim to remove any additional contributions (stellar photosphere, the accretion shocks, dusty disc, etc...). Following \citet{Kraus08, Bouvier20a}, we compute the continuum-subtracted observables, the so-called pure line visibility and phase, by using the emission line profiles computed in Sect. \ref{sect:MA_flow}. This is of prime importance in the case of \brg{} line as the magnetospheric emission is quite faint in the infrared ($\approx1.3$ excess flux compared to the continuum, Fig. \ref{fig:brg_allazim}). The derivation of the pure line quantities is only possible if the source is marginally resolved (i.e, size $< \lambda/2B$).

In this case, the pure line visibility $V_{line}$ and phase $\Phi_{line}$ are given by:
\begin{equation}
    V_{Line} = \frac{F_{L/C}V_{Tot} - V_{Cont}}{F_{L/C}-1},
    \label{eq:vline}
\end{equation}
\begin{equation}
    \Phi_{Line} = arcsin\left(\frac{F_{L/C}}{F_{L/C}-1}\frac{V_{Tot}}{V_{Line}}\sin{\Phi_{Tot}}\right).
\end{equation}
Where $F_{L/C}$ denotes the line-to-continuum flux ratio as taken from the normalised spectrum (Fig. \ref{fig:brg_allazim}), $V_{Cont}$ is the visibility computed in the continuum (star+shock only), and $V_{Tot}$, $\Phi_{Tot}$ are the total complex quantities measured by GRAVITY. In Eq.~(\ref{eq:vline}), we note that in the case when $F_{L/C}$ is close to one, the derived $V_{Line}$ cannot exist (converges to infinity). Such non-ideal profiles appear if the red absorption becomes too important. Therefore, we assume to discard the affected spectral channels for phases 0.25 and 0.47, where $F_{L/C}$ is too close to one: -- one point ($v=53$ \kms{}) at phases 0.25 and -- two points ($v=15$, 53 \kms{}) at phase 0.47.

\begin{figure}[h!]
    \centering
    \includegraphics[width=.85\columnwidth]{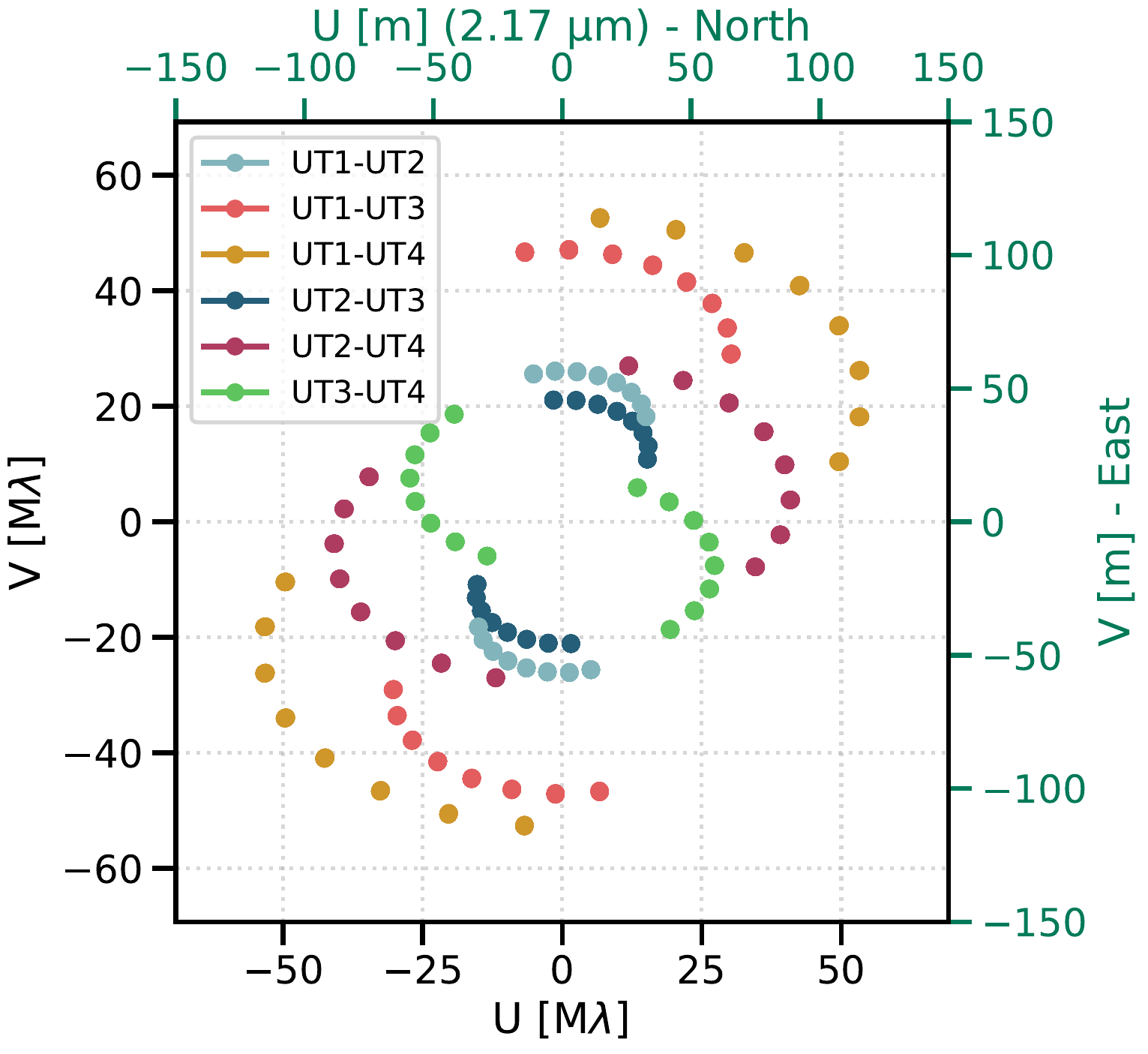}
    \caption{Fourier sampling (u-v coverage) of the simulated data. The different colours correspond to the six different baselines of the VLTI. The eight points per baseline represent a typical observational sequence with one data point per hour.}
    \label{fig:uv_coverage}
 \end{figure}

\begin{acknowledgements}
The authors thank Claudio Zanni, Lucas Labadie, Catherine Dougados, and Alexander Wojtczak for fruitful discussions.

This project has received funding from the European Research Council (ERC) under the European Union’s Horizon 2020 research and innovation programme (grant agreement No 742095; {\it SPIDI}: Star-Planets-Inner Disk-Interactions, \url{http://www.spidi-eu.org}). 

B. Tessore thanks the french minister of Europe and foreign affairs and the minister of superior education, research and innovation (MEAE and MESRI) for research funding through FASIC partnership.

C. Pinte acknowledges funding from the Australian Research Council via FT170100040 and DP180104235.

The numerical simulations presented in this paper were performed with the Dahu supercomputer of the GRICAD infrastructure (https://gricad.univ-grenoble-alpes.fr), which is supported by Grenoble research communities.
\end{acknowledgements}

\clearpage 
\bibliographystyle{aa}
\bibliography{bt_spidi2.bib}

\end{document}